\documentstyle[preprint]{jpsj}
\def\runtitle{Electronic states of doped spin ladders (Sr,Ca)$_{14}$Cu$_{24}$O$_{41}$}
\def\runauthor{Yoshiaki {\sc Mizuno}, Takami {\sc Tohyama} and Sadamichi {\sc Maekawa}}
\title{Electronic States of doped spin ladders (Sr,Ca)$_{14}$Cu$_{24}$O$_{41}$}
\author{Yoshiaki {\sc Mizuno}$^1$, Takami {\sc Tohyama}$^2$ and Sadamichi {\sc Maekawa}$^{1,3}$}
\inst{$^1$Department of Applied Physics, Nagoya University, Nagoya 464-01\\$^2$Department of Physics, Faculty of Education, Mie University, Tsu 514\\$^3$Institute for Materials Research, Tohoku University, Sendai 980-77}
\recdate{\today}

\abst{
We examine the electronic states of Sr$_{14-x}$Ca$_x$Cu$_{24}$O$_{41}$ by the ionic and cluster model approach.  It is found that self-doped holes are likely to stay on the chain at $x=0$ and the Ca substitution drives the holes to move to the ladder.  This feature is caused by the change of the positions of (Sr,Ca) layers, which enhances the electrostatic potentials in the chain.  The optical conductivity is calculated applying the exact diagonalization method to small Cu-O clusters.  It is shown that the excitations below 2~eV are governed by the ladder, giving rise to the spectral weight transfer from the charge-transfer excitation around 2~eV to low energy Drude excitation, while the contribution from the chain mainly emerges in higher energy region showing large weight around 3~eV.  These provide consistent explanation to recent experiment.}

\kword{Sr$_{14-x}$Ca$_x$Cu$_{24}$O$_{41}$, Madelung potential, optical conductivity,  exact diagonalization}

\begin{document}
\sloppy
\maketitle

The physical properties of (Sr,Ca)$_{14}$Cu$_{24}$O$_{41}$ have drawn much attention in connection with the recent discovery of superconductivity in Sr$_{0.4}$Ca$_{13.6}$Cu$_{24}$O$_{41}$ under high pressure.~\cite{Uehara} Sr$_{14-x}$Ca$_x$Cu$_{24}$O$_{41}$ contains Cu$_2$O$_3$ two-leg ladders as well as CuO$_2$ chains in which the CuO$_4$ units share their edges. Since six holes are left in the unit cell, the compounds are self-doped systems with the Cu average valence of +2.25.  The Ca substitution gives rise to remarkable changes in the transport and magnetic properties. The resistance for $x=0$ is semi-conductive but the substitution of Ca for Sr increases the conductivity.~\cite{Kato,Carter}  The spin gap existing in the ladder decreases with $x$, while the gap in the chain hardly changes.~\cite{Tsuji,Matsumoto}  These features are indicative of the redistribution of holes with the Ca substitution.  Three scenarios for the hole distribution have been proposed through the analyses of experimental results. (i) Most of holes are always on the chain.~\cite{Carter}  (ii) Holes are mainly on the ladder and independent of $x$.~\cite{Matsumoto}  (iii) For $x=0$, most of holes are on the chain.  With increasing $x$ the holes transfer from the chain to the ladder.~\cite{Kato,Tsuji,Osafune}

  In order to understand the physical properties of Sr$_{14-x}$Ca$_x$Cu$_{24}$O$_{41}$, it is crucial to clarify the distribution of the self-doped holes as a function of $x$.  In this letter, we examine the electronic states and the excitation properties of the compounds by means of the ionic and cluster model approach~\cite{Ohta1,Ohta2,Tohyama1} and clarify the hole distribution from a theoretical side.  We find that at $x=0$ the electrostatic potentials favor a situation that holes are on the chain, while with increasing $x$ the situation becomes unstable and the ladder is of advantage to the hole distribution.  This indicates the transfer of holes from the chain to the ladder with $x$.  By using exact diagonalization method, we calculate the optical conductivity $\sigma(\omega)$ for small clusters simulating the ladder and the chain.  When there are no holes in the ladder, i.e. for the case of $x$=0, the ladder shows the charge-transfer (CT) gap of 1.7~eV.  Upon hole doping, the spectral weights of the CT excitation transfer to the low-energy Drude contribution.  On the other hand, $\sigma(\omega)$ for the chain containing almost all of the holes shows high-energy structure around 2.6~eV.  This is in good accord with recent experiment of the optical conductivity.~\cite{Osafune}  Our theoretical results support the scenario (iii) mentioned above.

In order to investigate the distribution of the self-doped holes, we firstly perform the calculation of the Madelung potentials.  The crystal structure is taken from the data by McCarron {\it et al.}~\cite{McCarron} for $x=0$ and 8.  For other Ca content, we use extrapolated values of position parameters which are obtained by taking into account the variation of the lattice constants with $x$~\cite{Akimitsu}.  Table~I shows the total Madelung energy for various cases of the distribution of the 6 self-doped holes.  In the calculation, the holes are uniformly distributed on the oxygen in the ladder and/or the chain.  At $x=0$, the energy becomes minimum when all of holes are on the chain.  With increasing $x$, the energy minimum appears for the case where some of holes are on the ladder.  At $x=14$, the total Madelung energy favors a situation of no hole in the chain.  This result implies a tendency that the holes would move from the chain to the ladder with increasing Ca content.  

As another quantity that gives an information on the hole distribution, we consider the Madelung potential for a hole at the oxygen sites on the chain and the ladder, $V_{\rm oxygen}$.  This quantity is a measure of the distribution of an additional hole which is brought from outside.  In the calculation of $V_{\rm oxygen}$, the holes already existing in the system are uniformly distributed on Sr and/or Ca ions.~\cite{hole}  The results are shown as a function of $x$ in Fig.1.  The solid and dotted lines denote the potentials for the chain and the ladder, respectively.  At $x=0$,  $V_{\rm oxygen}$ for the chain is smaller than that for the ladder.   With increasing $x$, $V_{\rm oxygen}$ for the chain increases more rapidly than that for the ladder.  This rapid increase means that the oxygen on the chain becomes less favorable to an added hole with the substitution of Ca for Sr.  Therefore, the ladder reveals a tendency to occupy the hole.  This is consistent with the tendency of the total Madelung energy as shown in Table~I.  

Such remarkable features originate from the variation of positional parameters by the Ca substitution; the (Sr,Ca) layers which have positive charges approach closer to the chain layers than to the ladder ones, resulting in the rapid enhancement of the potential for a hole in the chain.  

Note that the above results obtained from the Madelung energy calculations  show only the tendency of hole distribution and can not tell us the correct number of distributed holes, because covalency which is important to the cuprates is not included.  For example, one can see from Fig.~1 that $V_{\rm oxygen}$ for the chain is lower than that for the ladder, even if all of Sr is replaced by Ca ($x=14$).  Thus, one might argue that all of the holes are always on the chain.  This argument, however, is not correct because of the lack of the effect of covalency.  Actually, the covalency has a tendency to make the energy of hole higher in the chain than in the ladder because of the difference of the geometry for Cu-O bonding (180 degree Cu-O bonding for the ladder and nearly 90 degree Cu-O bonding for the chain).

Next, we examine the optical conductivity $\sigma(\omega)$ by taking the covalency into account. The present compounds are classified as a charge-transfer (CT) type. Important quantities governing their electronic states are the energy-level separation between the Cu3$d$ and O2$p$ orbitals, i.e. the CT energy $\Delta$, the hopping energy between the two orbitals and between O2$p$ orbitals, and on-site Coulomb interactions on Cu and O.  $\Delta$ may be written as~\cite{Ohta1} $\Delta = \Delta V_M / \epsilon_\infty + \Delta_0$, where $\Delta V_M$ is the difference in the Madelung potential for a hole between copper and oxygen sites, $\epsilon_\infty$ is the dielectric constant, and $\Delta_0$ is related to the second ionization energy of a Cu$^{2+}$ ion and the second electron affinity of an O$^{2-}$ ion.  $\Delta V_M$ is evaluated from the Madelung potential calculations mentioned above.  $\Delta_0$ is taken to be $-$10.88~eV as for other high $T_{\it c}$ cuprates.~\cite{Ohta1}  $\epsilon_\infty$ is set to be a rounded value of 3.4 between the values of 3.5 for pure ladder systems~\cite{Mizuno} (and plane systems~\cite{Ohta1}) and of 3.3 for chain systems,~\cite{Tohyama2} because the present compound contains both the ladder and chain.  The hopping energies are determined according to the previous study on the high-$T_c$ cuprates~\cite{Tohyama3}.  The bond length dependence with powers of $-$4 and $-$3 is assumed for the hopping parameters between 3$d$ and 2$p$ orbitals and between 2$p$ orbitals, respectively.~\cite{pds}  The parameters used are listed in Table~II.~\cite{U}

The exact diagonalization method is used for the calculation of $\sigma(\omega)$ for small clusters.~\cite{Tohyama4,Tohyama5}  We employ a Cu$_6$O$_{17}$ cluster and a Cu$_4$O$_{10}$ cluster with open boundary condition for the ladder and the chain, respectively.~\cite{cluster}  The former cluster has three Cu sites per leg.  The CuO$_4$ units share their corners along the leg and the rung.  The latter one contains four CuO$_4$ units sharing their edges to form Cu-O chain. 

We first show the results for the ladder.  From the investigation of the electrostatic potentials, it was found that self-doped holes are likely to stay on the chain at $x=0$, and the Ca substitution lets the holes move to the ladder.  Therefore, we calculate $\sigma(\omega)$ for the undoped and one-hole doped Cu$_6$O$_{17}$ clusters.  The results are shown in Fig.2.  In Fig.2(a), the lowest energy structure gives the CT gap of 1.7~eV.  In the schematic description of the density of states (DOS) depicted in Fig.3, the CT excitation is identified as the excitation from the Zhang-Rice local singlet band~\cite{Zhang} to the upper Hubbard band denoted by the process {\bf A}.  Upon hole doping, the spectral weight of the CT excitation moves to the low energy Drude weight around $\omega$=0.4~eV as shown in Fig.2(b).  Such spectral weight transfer is common to all cuprates containing  corner-sharing CuO$_4$ units.~\cite{Tohyama4,Wagner}  The large Drude weight indicates that the ladder becomes metallic.  

For $x$=0, almost all of the self-doped holes are on the chain, i.e. six holes enter ten CuO$_2$ units in the chain per unit cell.  Such a situation can be simulated by the two-hole doped Cu$_4$O$_{10}$ cluster.  Shown in Fig.4 are the results for the two-hole doped case together with the undoped case for comparison.~\cite{LaCa}  In the undoped case, there are two regions where spectral weight is concentrated: one is located around 2~eV and the other around 4.2~eV.  The former region is due to the CT excitation.  Its intensity is small compared to the case of the ladder, because the hole propagation to neighboring CuO$_4$ units is suppressed due to the near 90 degree (93.4 degree in case of  Sr$_{8}$Ca$_{6}$Cu$_{24}$O$_{41}$) edge-sharing Cu-O bonding.  This is a direct manifestation of the structural difference between edge-sharing and corner-sharing CuO$_4$ units.  The latter region is related to a local excitation within CuO$_4$ unit.  In the schematic description of DOS depicted in Fig.3(a), the excitation is represented by {\bf B}, i.e. from the O2$p$ non-bonding bands to the upper Hubbard bands.

When the holes are introduced(Fig.4(b)), a new structure with strong intensity appears  around 2.6~eV.  The structure comes from excitations from O2$p$ non-bonding bands to unoccupied Zhang-Rice local single bands, denoted by {\bf C} in Fig.3(b).  This is also a local excitation in the sense that a single CuO$_4$ cluster is enough to see such a structure with strong intensity.  In other words, the excitation has nothing to do with the hole propagation.  The strong intensity in Fig.4(b) indicates that doped holes are nearly localized by the edge-sharing Cu-O bond.  The intensity is strongly dependent on the hole concentration and reduced by half in one-hole doped case(25$\%$ doping). Therefore the structure may be a measure of the hole distribution.  Note that the corresponding structure in the ladder overlaps with the CT excitation.  Small amount of delocalized holes produce the low-energy Drude weight at $\omega$=0.5~eV, which may  contribute to the metallic behavior.  However, such small conductivity will be taken away by the effect of the localization due to low dimensionality.  

From these results, we conclude that $\sigma(\omega)$ below 2~eV are dominated by the ladder while the contribution from the chain mainly emerges in higher energy region above 2~eV.  Recent experiments by Osafune {\it et al.}~\cite{Osafune} show the following features: (i) The spectral weight around 2~eV for $x=0$ decreases with increasing $x$ and simultaneously the Drude weight increases.  (ii) There is a large spectral weight around 2.8~eV and it slightly decreases with $x$.  The feature (i) is in remarkable agreement with our results of the ladder.  The weight in (ii) can be attributed to the excitation of the chain at 2.6~eV in Fig.4(b).  The slight decrease of the weight with $x$ indicates that the decrease of the number of holes in the chain is small. This means small number of holes in the ladder and justifies our simulation of the ladder with one hole in the six Cu sites.  The good agreement with the experiments strongly confirm our arguments that the Ca substitution enforces the transfer of holes from the chain to the ladder. 

  In summary, we have examined the electronic states and excitation properties of Sr$_{14-x}$Ca$_x$Cu$_{24}$O$_{41}$ by the ionic and cluster model approach.  It is found that self-doped holes are likely to stay on the chain at $x=0$ and the Ca substitution drives the holes to move to the ladder.  This remarkable feature of the compound is caused by the change of the positions of (Sr,Ca) layers, which enhances the electrostatic potentials in the chain more strongly than in the ladder.  The recent experimental results of the optical conductivity are consistent with our theoretical prediction: The excitations below 2~eV are governed by the ladder, giving rise to the spectral weight transfer from the CT excitation around 1.7~eV to low energy Drude excitation, while the contribution from the chain mainly emerges in higher energy region showing large weight around 2.6~eV.

We would like to thank J. Akimitsu, S. Uchida and H. Eisaki for informing us their experimental data prior publication. This work was supported by a Grant-in-Aid for Scientific Research on Priority-Areas from the Ministry of Education, Science, Sports and Culture of Japan. We thank the Supercomputer Center, Inst. for Solid State Phys., Univ. of Tokyo, Japan for allocation of CPU time on the FACOM VPP500 supercomputer. The author(Y.M.)  acknowledges the financial support by JSPS Research Fellowships for Young Scientists.

\newpage
\center{REFERENCES}

\newpage
\center{TABLES}
\begin{table}
\caption{The dependence of the total Madelung energy of Sr$_{14-x}$Ca$_x$Cu$_{24}$O$_{41}$ on the distribution of self-doped holes.}
\label{table:1}
\begin{tabular}{@{\hspace{\tabcolsep}\extracolsep{\fill}}ccccc} \hline
\multicolumn{2}{c}{The number of holes} & \multicolumn{3}{c}{Madelung energy (eV)}\\ 
chain & ladder & $x$=0 & $x$=8 & $x$=14\\ \hline
 6 & 0  & -1764	& -1792 & -1824\\
 3 & 3  & -1762 & -1794 & -1830\\
 0 & 6  & -1757 & -1789 & -1831\\ \hline
\end{tabular}
\end{table}

\begin{table}
\caption{The parameters used in the cluster calculation. Listed are the bond length between Cu and O ($d_{\rm Cu-O}$ (\AA) ), the charge-transfer energy ($\Delta$ (eV) ) and the Slater-Koster hopping parameters (energy unit in eV) for a hole.}
\label{table:2}
\begin{tabular}{@{\hspace{\tabcolsep}\extracolsep{\fill}}ccc} \hline
 & chain & ladder\\ \hline
$d_{\rm Cu-O}$ & 1.89 & 1.96$^a$, 1.88$^b$\\ 
$\Delta$ & 48.15 & 47.85$^a$, 47.80$^b$\\
($pd\sigma$) & 1.342 & 1.172$^a$, 1.348$^b$\\
($pd\pi$) & -0.619 & -0.540$^a$, -0.621$^b$\\
($pp\sigma$) & -0.658$^c$, -0.789$^d$ & -0.684\\
($pp\pi$) & 0.165$^c$, 0.197$^d$& 0.171\\ \hline
\end{tabular}
$^a$leg direction, $^b$rung direction, $^c$parallel direction to the chain, $^d$perpendicular direction to the chain
\end{table}

\newpage
\center{FIGURE CAPTIONS}
\begin{figure}
\caption{The Ca content ($x$) dependence of the Madelung potentials (V$_{\rm{oxygen}}$(eV)) at the chain-oxygen and ladder-oxygen sites.} 
\label{fig:1}
\end{figure}
\begin{figure}
\caption{Optical conductivity $\sigma(\omega)$ for the Cu$_{6}$O$_{17}$ cluster simulating the ladder. (a) undoped case, (b) one-hole doped case (16.7\% doping).  Here, $d_{\rm Cu-O}$ and $e$ are the bond length to leg direction between Cu and O and the elementary electric charge,  respectively, and $\hbar = c = 1$. The $\delta$ functions are convoluted with a Lorentzian broadening of 0.02~eV. The labels {\bf A} and {\bf B} denote the processes of the interband excitations depicted in Fig. 3.}
\label{fig:2}
\end{figure}
\begin{figure}
\caption{The schematic description of the density of states for cuprates.  (a) undoped case, (b) hole-doped case.  LH: lower Hubbard band, UH: upperHubbard band, NB: non-bonding 2$p$ band, Z-R: Zhang-Rice local singlet band.  The labels {\bf A},{\bf B} and {\bf C}, denote the processes of optical interband excitations.}
\label{fig:3}
\end{figure}
\begin{figure}
\caption{Optical conductivity $\sigma(\omega)$ for the Cu$_{4}$O$_{10}$ cluster simulating the edge-sharing CuO$_4$ chain.  (a) undoped case, (b) two-hole doped case (50\% doping). Here, $d_{\rm Cu-O}$ and $e$ are the bond length between Cu and O and the elementary electric charge,  respectively, and $\hbar = c = 1$. The $\delta$ functions are convoluted with a Lorentzian broadening of 0.02~eV. The labels {\bf A}, {\bf B} and {\bf C} denote the processes of the interband excitations depicted in Fig. 3.}
\label{fig:4}
\end{figure}


\begin{thebibliography}{99}
\bibitem{Uehara} M. Uehara, T. Nagata, J. Akimitsu, H. Takahashi, N. Mori and K. Kinoshita: J. Phys. Soc. Jpn. {\bf 65} (1996) 2764.
\bibitem{Kato} M. Kato, K. Shiota and Y. Koike: Physica C {\bf 258} (1996) 284.
\bibitem{Carter} S. A. Carter, B. Batlogg, R. J. Cava, J. J. Krajewski, W. F. Peck, Jr. and T. M. Rice: Phys. Rev. Lett. {\bf 77} (1996) 1378.
\bibitem{Tsuji} S. Tsuji, K. Kumagai, M. Kato and Y. Koike: J. Phys. Soc. Jpn. {\bf 65} (1996) 3474.
\bibitem{Matsumoto} S. Matsumoto, Y. Kitaoka, K. Magishi, K. Ishida and K. Asayama: preprint.
\bibitem{Osafune} T. Osafune, N. Motoyama, H. Eisaki and S. Uchida: submitted to Phys. Rev. Lett.
\bibitem{Ohta1} Y. Ohta, T. Tohyama and S. Maekawa: Phys. Rev. B {\bf 43} (1991)  2968.
\bibitem{Ohta2} Y. Ohta, T. Tohyama and S. Maekawa: Phys. Rev. Lett. {\bf 66} (1991) 1228.
\bibitem{Tohyama1} T. Tohyama and S. Maekawa: J. Phys. Soc. Jpn. {\bf 65} (1996) 667.
\bibitem{McCarron} E. M. McCarron, III, M. A. Subramanian, J. C. Calabrese and R. L. Harlow: Mat. Res. Bull. {\bf 23} (1988) 1355.
\bibitem{Akimitsu} J. Akimitsu: private communication.
\bibitem{hole} This situation is equivalent to considering La$_6$Ca$_8$Cu$_{24}$O$_{41}$ which has no holes on both the ladder and the chain, i.e. the parent compound of the present system.  The conclusions obtained here do not depend on a way of distributing holes.
\bibitem{Mizuno} Y. Mizuno, T. Tohyama and S. Maekawa: unpublished.
\bibitem{Tohyama2} T. Tohyama, Y. Mizuno, S. Maekawa, C. Kim and Z.-X. Shen: to be published in Physica B.
\bibitem{Tohyama3} T. Tohyama and S. Maekawa: J. Phys. Soc. Jpn. {\bf 59} (1990) 1760.
\bibitem{pds} ($pd\sigma$)=1.3~eV at $d_{\rm Cu-O}$=1.905~\AA~is used as a standard hopping parameter.  The value is taken from ref.~7).
\bibitem{U} The on-site Coulomb energies are set to be $U_d$=8.5~eV and $U_p$=4.1~eV for Cu and O, respectively.  The Hund's coupling of 0.6~eV on oxygen is included in the cluster calculation of the chain.
\bibitem{Tohyama4} T. Tohyama and S. Maekawa: J. Phys. Soc. Jpn. {\bf 60} (1991) 53.
\bibitem{Tohyama5} T. Tohyama and S. Maekawa: Physica C {\bf 191} (1992) 193. 
\bibitem{cluster} In the clusters, we take single Cu3$d$ orbital ($d_{x^2-y^2}$). As for  oxygen sites, single and double O2$p$ orbitals are used for the ladder and the chain, respectively. 
\bibitem{Zhang} F. C. Zhang and T. M. Rice: Phys. Rev. B {\bf 37} (1988) 3759.
\bibitem{Wagner} J. Wagner, W. Hanke and D. J. Scalapino: Phys. Rev. B {\bf 43} (1991) 10517.
\bibitem{LaCa} The half-filled chain is realized in a compound La$_6$Ca$_8$Cu$_{24}$O$_{41}$.

\end{thebibliography}
\end{document}